\documentclass[preprint]{jpsj2}

\def\mathvc#1{\mbox{\boldmath $\mathit{#1}$}}

\title{Theory of Superconductivity in PuCoGa$_{5}$}

\author{Kazunori \textsc{Tanaka} \thanks{E-mail address: tanaka.kazunori@scphys.kyoto-u.ac.jp},      Hiroaki \textsc{Ikeda} and Kosaku \textsc{Yamada}}

\inst{Department of science,      Kyoto university,      Kyoto 606-8502,      Japan}
\recdate{July 31, 2003}

\abst{Recently,      superconductivity in PuCoGa$_{5}$ was discovered. It has the same crystal structure as Ce$M$In$_{5}$($M=$Ir,     Co,     Rh),      which are often refered to as Ce-115's. The electron correlation in PuCoGa$_{5}$ is estimated to be weak compared with Ce-115's,      and the filling number of electrons is considered to be far from 0.5/spin in the band which plays an important role in realizing the superconductivity. Nevertheless,     the superconducting transition temperature $\mathit{T_{c}}$ in PuCoGa$_{5}$ is almost by an order of magnitude  higher than that in Ce-115's. In order to explain the superconductivity with high $\mathit{T_{c}}$,      we adopt the periodic Anderson model and calculate $\mathit{T_{c}}$ by solving the Dyson-Gor'kov equation derived by the third order perturbation theory with respect to $U$. By this calculation,      we indicate  that the superconducting state of PuCoGa$_{5}$ is  a $d$-wave pairing state,      and show that the good location of two Fermi surfaces results in the high $\mathit{T_{c}}$ in PuCoGa$_{5}$.}

\kword{unconventional superconductivity,      periodic Anderson model,      heavy Fermion,   plutonium}

\begin{document}
\maketitle
\section{Introduction}

Recently,      superconductivity in PuCoGa$_{5}$ was discovered by Sarrao,     \textit{et al.\ }\cite{bib1} It has very high transition temperature ($\mathit{T_{c}}=18.5\mathrm{K}$). This  value of $\mathit{T_{c}}$ is higher than that in any other isostructual superconductors,      such as Ce$M$In$_{5}$ ($M=$Ir,     Co,     Rh). Superconducting transition in CeCoIn$_5$ and CeIrIn$_5$ occur at ambient pressure at $\mathit{T_{c}}=2.3\mathrm{K}\  \mathrm{and} \ 0.4\mathrm{K}$,      respectively. \cite{bib2,     bib3} But CeRhIn$_5$ becomes superconducting only under pressure with $\mathit{T_{c}}=2.1\mathrm{K}$. \cite{bib4} Although NpCoGa$_{5}$ and U$M$Ga$_{5}$ have the same HoCoGa$_{5}$-type crystal structure,      superconductivity has  never been reported in these materials. From now on,      we refer to these HoCoGa$_{5}$-type compounds as `115'.

First of all,      let us consider Ce-115's. In the phase diagram of Ce-115's,      antiferromagnetic(AF) state and superconducting  state are adjacent to each other. \cite{bib5} Moreover,      the magnetic field dependence of thermal conductivity \cite{bib6} and $T^3$-behavior of nuclear spin relaxation rate in the superconducting state \cite{bib7} show the existence of line-node gap. Band calculation  shows that the Fermi surfaces of Ce-115's are quasi-two-dimensional. From these facts,   Ce-115's have been considered to be unconventional quasi-two-dimensional \textit{d}-wave superconductors induced by antiferromagnetic spin fluctuations (AFF).\cite{bib8} Using Hubbard model,      Nisikawa \textit{et al.\ } explained that the superconductivity of Ce-115 has $d_\mathrm {x^{2}-y^{2}}$ symmetry.\cite{bib9}

 Then,      what is the mechanism of superconductivity in PuCoGa$_{5}$? Experimental facts such as the Curie-Weiss behavior in magnetic susceptibility at $T>\mathit{T_{c}}$,      $T^{1.35}$-behavior in electric resistivity at  $\mathit{T_{c}}<T<50\mathrm{K}$,      and power-law behavior in specific heat at $T<\mathit{T_{c}}$  are reported.\cite{bib1} Moreover, band calculations by Maehira \textit{et al.\ }\cite{bib10} and Opahle \textit{et al.\ }\cite{bib11} show that  the Fermi surfaces of PuCoGa$_{5}$ are quasi-two-dimensional just like Ce-115's. These facts imply that the superconductivity in PuCoGa$_{5}$ is an unconventional $d$-wave superconductivity with magnetic origin,      just like Ce-115's. Of course,      there are differences between PuCoGa$_{5}$ and Ce-115's. One is the value of $\mathit{T_{c}}$,      and another is the strength of electron correlation. Specific heat coefficients $\gamma  =  \left. {C/T} \right|_{T = T_c } $,      which are  proportional to the renormalized electron mass   $m^*$,      are  290,      400 and    750 ${\rm{mJ/mol}} \cdot {\rm{K}}^{\rm{2}} $ for CeCoIn$_5$,      CeRhIn$_5$ and CeIrIn$_5$,     respectively.\cite{bib4} On the other hand,      $\gamma$ for PuCoGa$_5$ is 77${\rm{mJ/mol}} \cdot {\rm{K}}^{\rm{2}} $. This means that the mass of electron is not so enhanced and the electron correlation is modest in PuCoGa$_5$ compared with Ce-115's. The theoretical specific heat coefficient $\gamma _{{\rm{band}}} $   estimated from the band calculation \cite{bib10}  is 16.9${\rm{mJ/mol}} \cdot {\rm{K}}^{\rm{2}} $. Thus we can estimate the mass enhancement factor $m^* /m = \gamma /\gamma_{{\rm{band}}} $ in PuCoGa$_5$ at $ 4.5$. This value is rather lower than that in Ce-115's,   which is more than 10.

 From Ref. \citen{bib10} we can see  that    in PuCoGa$_{5}$ there exist no bands which are near the half-filled. From this fact, it seems difficult to explain $\mathit{T_{c}}$   in PuCoGa$_{5}$  which is high almost by an order of magnitude  compared with Ce-115's. Let us see the Fermi surfaces of PuCoGa$_5$ in band calculations. \cite{bib10,     bib11} The 16th band and the 17th band in Ref.\citen{bib10} have   Fermi surfaces. Here we ignore the Fermi surfaces of the 15th band and the 18th band since they are very small. In these situations, the following points are important.  Since PuCoGa$_5$ has the two main  Fermi surfaces,      the effective density of state at the Fermi energy becomes large. Furthermore,   if two Fermi surfaces are well located,  the effective correlation for antiferromagnetic fluctuation is strengthened as shown below,      even though $\gamma$ is still not so large. In this paper we point out that this leads to  relatively high $\mathit{T_{c}}$ for PuCoGa$_5$.

\section{\label{sec3} Periodic Anderson Model}
 Let us introduce the following periodic Anderson model\cite{bib12}. The Hamiltonian is
\begin{eqnarray}
H = \sum\limits_{{\mathvc{k}},\mathit{\sigma} } {\left[ {\varepsilon _{^{\mathvc{k}} }^f f_{{\mathvc{k}}\mathit{\sigma} }^\dag  f_{{\mathvc{k}}\mathit{\sigma} }  + \varepsilon _{^{\mathvc{k}} }^c c_{{\mathvc{k}}\mathit{\sigma} }^\dag  c_{{\mathvc{k}}\mathit{\sigma} }  + V_{\mathvc{k}} \left( {f_{{\mathvc{k}}\mathit{\sigma} }^\dag  c_{{\mathvc{k}}\mathit{\sigma} }  + c_{{\mathvc{k}}\mathit{\sigma} }^\dag  f_{{\mathvc{k}}\mathit{\sigma} } } \right)} \right]} \nonumber \\
{}+ \frac{U}
{N}\sum\limits_{{\mathvc{k}},{\mathvc{k'}}} {f_{{\mathvc{k}} \uparrow }^\dag  f_{{\mathvc{q}} - {\mathvc{k}} \downarrow } f_{{\mathvc{q - k'}} \uparrow }^\dag  f_{{\mathvc{k'}} \downarrow } },   \label{eq1} \\
\varepsilon _{\mathvc{k}}^f  = 2t(\cos k_x  + \cos k_y ) + 4t' \cos k_x \cos k_y  - \mu _0,   \label{eq2} \\
\varepsilon _{\mathvc{k}}^c  = 2t_c (\cos k_x  + \cos k_y ) + 4t_c' \cos k_x \cos k_y + \mu _c - \mu _0,      \label{eq3} \\
V_{\mathvc{k}}  = V_0  - V_1 \cos k_x \cos k_y. \label{eq4} 
\end{eqnarray}
%
Here, $t$ and $t'$ denote the nearest and next nearest neighbor hopping terms of $f$-electron, respectively. $t_c$ and $t_c'$ denote those of conduction electron.  Thus, $ \varepsilon _{\mathvc{k}}^f $ and $\varepsilon _{\mathvc{k}}^c$ are the dispersion of $f$-electron and conduction-electron,      respectively. $ V_{\mathvc{k}}  $ is the hybridization between $f$-electron and conduction-electron.  We set these parameters so that the band structure and the Fermi surfaces of  diagonalized bands reproduce those of the  band calculation in Ref.\citen{bib10} and Ref.\citen{bib11}. Hereafter we use the following parameters: $\ t_c/t  = 6.0$,  $\ t_c'/t  = 1.8$,          $V_0/t  = 2.8$,      $V_1/t  = 2.1$,  $\mu _c/t=0.8$   and $t'/t  = 0.3$. The total number of filled electrons $n_{tot}$ in the  $f$-band and the conduction-band is 1.16 per spin (58\% filled). The chemical potential $\mu _0$ at temperature $T$ is determined by 
%
\begin{eqnarray}
\frac{1}
{N}\sum\limits_{\mathvc{k}} {\left( {f\left( {\varepsilon _{\mathvc{k}}^f } \right) + f\left( {\varepsilon _{\mathvc{k}}^c } \right)} \right) = n_{tot} }, \label{eq101}
\end{eqnarray}
%
where $f\left( x  \right) = \left( {e^x  + 1} \right)^{ - 1} $ is the Fermi distribution function. The unperturbed term of the Hamiltonian of Eq.(\ref{eq1}) is rewritten in the $2 \times 2$ matrix form as follows
\begin{eqnarray}
H_0 = \left( {\begin{array}{*{20}c}
   {f_{{\mathvc{k}}\mathit{\sigma} }^\dag  } & {c_{{\mathvc{k}}\mathit{\sigma} }^\dag  }  \\
\end{array}} \right)\left( {\begin{array}{*{20}c}
   {\varepsilon _{\mathvc{k}}^f } & {V_{\mathvc{k}} }  \\
   {V_{\mathvc{k}} } & {\varepsilon _{\mathvc{k}}^c }  \\
\end{array}} \right)\left( {\begin{array}{*{20}c}
   {f_{{\mathvc{k}}\mathit{\sigma} } }  \\
   {c_{{\mathvc{k}}\mathit{\sigma} } }  \\
\end{array}} \right) \nonumber \\
{}  = \left( {\begin{array}{*{20}c}
   {f_{{\mathvc{k}}\mathit{\sigma} }^\dag  } & {c_{{\mathvc{k}}\mathit{\sigma} }^\dag  }  \\
\end{array}} \right)\left( {\begin{array}{*{20}c}
   c & { - s}  \\
   s & c  \\
\end{array}} \right)\left( {\begin{array}{*{20}c}
   {E_1 } & 0  \\
   0 & {E_2 }  \\
\end{array}} \right)\left( {\begin{array}{*{20}c}
   c & s  \\
   { - s} & c  \\
\end{array}} \right)\left( {\begin{array}{*{20}c}
   {f_{{\mathvc{k}}\mathit{\sigma} } }  \\
   {c_{{\mathvc{k}}\mathit{\sigma} } }  \\
\end{array}} \right) \nonumber \\
{}  = \left( {\begin{array}{*{20}c}
   {a_{{\mathvc{k}}\mathit{\sigma} }^\dag  } & {b_{{\mathvc{k}}\mathit{\sigma} }^\dag  }  \\
\end{array}} \right)\left( {\begin{array}{*{20}c}
   {E_1 } & 0  \\
   0 & {E_2 }  \\
\end{array}} \right)\left( {\begin{array}{*{20}c}
   {a_{{\mathvc{k}}\mathit{\sigma} } }  \\
   {b_{{\mathvc{k}}\mathit{\sigma} } }  \\
\end{array}} \right),      \label{eq5} \\
E_{\scriptstyle 1 \hfill \atop 
  \scriptstyle 2 \hfill}  = \frac{1}{2}\left( {\left( {\varepsilon _{\mathvc{k}}^c  + \varepsilon _{\mathvc{k}}^f } \right) \pm \sqrt {\left( {\varepsilon _{\mathvc{k}}^f  - \varepsilon _{\mathvc{k}}^c } \right)^2  + 4V_{\mathvc{k}} ^2 } } \right). \label{eq6} 
\end{eqnarray}
Here,     the $f$-band and  the conduction-band are hybridized by  $V_{\mathvc{k}}  $,      and then diagonalized into two bands: the band-1 and the band-2. Operators $ f_{{\mathvc{k}}\mathit{\sigma} } \left( {f_{{\mathvc{k}}\mathit{\sigma} }^\dag  } \right)$,      $c_{{\mathvc{k}}\mathit{\sigma} } \left( {c_{{\mathvc{k}}\mathit{\sigma} }^\dag  } \right)$,      $ a_{{\mathvc{k}}\mathit{\sigma} } \left( {a_{{\mathvc{k}}\mathit{\sigma} }^\dag  } \right)$ and $b_{{\mathvc{k}}\mathit{\sigma} } \left( {b_{{\mathvc{k}}\mathit{\sigma} }^\dag  } \right)$ are the annihilation(creation) operators of the $f$-band,   the   conduction band,       the band-1 and the band-2,      respectively.  $ E_1 \;{\rm{and}}\;E_2 $ are the dispersions of the diagonalized bands. $ \left( {E_1  > E_2 } \right)$ 

We now  can express the bare Green's function as follows
\begin{eqnarray}
\hat G_0 ({k}) = \left( {\begin{array}{*{20}c}
   {G_0^f } & {G_0^{fc} }  \\
   {G_0^{cf} } & {G_0^c }  \\
\end{array}} \right)  \nonumber \\
{}  = \left( {\begin{array}{*{20}c}
   {c^2 G_1  + s^2 G_2 } & {sc(G_1  - G_2 )}  \\
   {sc(G_1  - G_2 )} & {s^2 G_1  + c^2 G_2 }  \\
\end{array}} \right),     \label{eq7} \\
G_{\scriptstyle 1 \hfill \atop 
  \scriptstyle 2 \hfill}  = \frac{1}{{i\varepsilon _n  - E_{\scriptstyle 1 \hfill \atop 
  \scriptstyle 2 \hfill} }},     \label{eq8} \\
sc = \frac{{V_{\mathvc{k}} }}{{\sqrt {\left( {\varepsilon _{\mathvc{k}}^f  - \varepsilon _{\mathvc{k}}^c } \right)^2  + 4V_{\mathvc{k}} ^2 } }} \label{eq9},     \\
c^2  = \frac{1}{2} + \frac{{\varepsilon _{\mathvc{k}}^f  - \varepsilon _{\mathvc{k}}^c }}{{2\sqrt {\left( {\varepsilon _{\mathvc{k}}^f  - \varepsilon _{\mathvc{k}}^c } \right)^2  + 4V_{\mathvc{k}} ^2 } }} \label{eq10},     \\
s^2  = 1 - c^2.   \label{eq11}
\end{eqnarray}  
In Eq.(\ref{eq8}),  $\varepsilon _n  = \left( {2n + 1} \right)\pi T$ is a fermion Matsubara frequency.

\section{\label{sec4} Calculation by TOPT}
In our numerical calculation,      we divide the first Brillouin zone into $ 128 \times 128$ meshes and take 4096 Matsubara frequencies. To  treat electron correlation effect,      we need to approximate the self-energy terms. Among several ways of approximation,      we adopt the third-order perturbation theory (TOPT) with respect to U. Using TOPT,      we can write the self-energy terms as
\begin{eqnarray}
\mathit{\Sigma} _n^f \left( k \right) = \frac{T}{N}\sum\limits_{k'} {V_{\rm{n}} \left( {k,   k'} \right)G_0^f \left( {k'} \right)},     \label{eq12} \\
V_{\text{n}} \left( {k,k'} \right) = \frac{T}
{N}U^2 \chi _{f0} \left( {k - k'} \right) + \frac{T}
{N}U^3 \left[ {\chi _{f0} ^2 \left( {k - k'} \right) + \phi _{f0} ^2 \left( {k + k'} \right)} \right],     \label{eq13} \\
\chi _{f0} \left( q \right) =  - \frac{T}{N}\sum\limits_k {G_0^f \left( k \right)} G_0^f \left( {k + q} \right),     \label{eq14} \\
\phi _{f0} \left( q \right) =  - \frac{T}{N}\sum\limits_k {G_0^f \left( k \right)} G_0^f \left( {q - k} \right).  \label{eq15} 
\end{eqnarray}
Here,  we have introduced the abbreviation $k = \left( {{\mathvc{k}}, \varepsilon _n } \right)$ and $q = \left( {{\mathvc{q}}, \omega _n } \right)$. Note that $\omega _{\rm{n}}  = 2n\pi T$ is a boson Matsubara frequency. 
The dressed Green's function $\hat G(k)$ in normal state is given by
\begin{eqnarray}
\hat G(k) = \hat G_0 (k) + \hat G_0 (k) \mathit{\hat \Sigma} \left( k \right)\hat G(k),  \label{eq16} 
\end{eqnarray}
where,     
\begin{eqnarray}
\hat G_0 (k) = \left( {\begin{array}{*{20}c}
   {G_0^f } & {G_0^{fc} }  \\
   {G_0^{cf} } & {G_0^c }  \\
\end{array}} \right) \label{eq17},      \\
\hat G(k) = \left( {\begin{array}{*{20}c}
   {G_{}^f } & {G_{}^{fc} }  \\
   {G_{}^{cf} } & {G_{}^c }  \\
\end{array}} \right) \label{eq18},     \\
\mathit{\hat \Sigma} \left( k \right) = \left( {\begin{array}{*{20}c}
   {\left( {\mathit{\Sigma} _n^f  - \delta \mu } \right)} & {\rm{0}}  \\
   {\rm{0}} & {\rm{0}}  \\
\end{array}} \right). \label{eq19}
\end{eqnarray}
The shift of chemical potential  ${\delta \mu }$ is determined by conservation of total electron number
\begin{eqnarray}
\sum\limits_k {\left( {G_{}^f  + G^c  - G_0^f  - G_0^c } \right)}  = 0.  \label{eq20}
\end{eqnarray}
It is noted that the self-energy correction  appears only in $f$-electrons, since the electron correlation $U$ is taken into account only among $f$-electrons.

Now,  we can calculate $E_1 ^\prime  $ and $E_2 ^\prime  $,   which are the modified dispersions of $E_1$ and $E_2$ by including the self-energy correction, respectively. They are given by 
%
%
\begin{eqnarray}
E_{\scriptstyle 1 \hfill \atop 
  \scriptstyle 2 \hfill} ^\prime   = \frac{1}{2}\left[ {\left( {\varepsilon _{\mathvc{k}}^c  + \varepsilon _{\mathvc{k}}^f  + \mathit{\Sigma} _n^R\left( {{\mathit{\mathvc{k}}},\varepsilon  = 0} \right)  - \delta \mu } \right) \pm \sqrt {\left( {\varepsilon _{\mathvc{k}}^f  + \mathit{\Sigma} _n^R\left( {{\mathit{\mathvc{k}}},\varepsilon  = 0} \right)  - \delta \mu  - \varepsilon _{\mathvc{k}}^c } \right)^2  + 4V_{\mathvc{k}} ^2 } } \right], \label{eq102}
\end{eqnarray}
%
%
\begin{figure}
\includegraphics[]{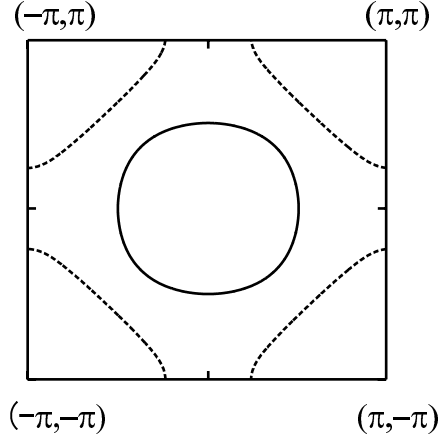}
\caption[]{The Fermi surfaces of the band-1(dashed curve) and the band-2(solid curve). }
\label{fig1}
\end{figure}
\begin{figure}
\includegraphics[]{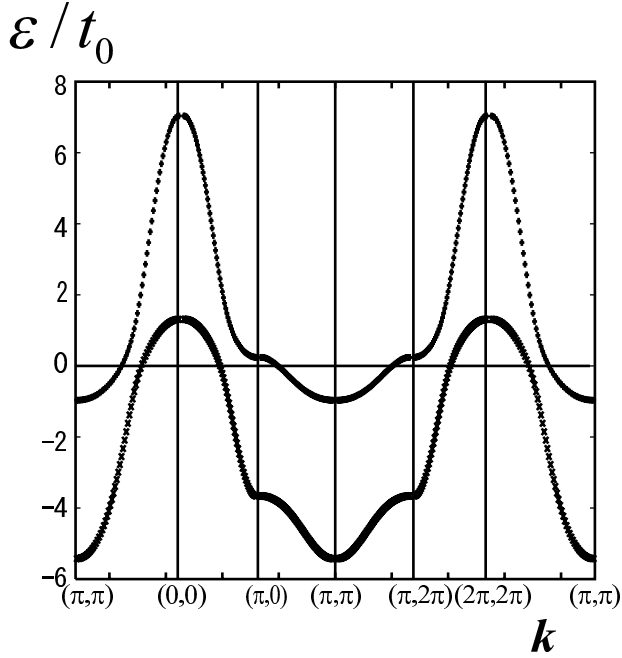}
\caption[]{The dispersion of the band-1(upper) and the band-2(lower). }
\label{fig2}
\end{figure}
%
%
where $\mathit{\Sigma} _n^R \left( {{\mathvc{k}}, \varepsilon } \right)$ is  the retarded normal self-energy,  which is calculated by analytic continuation ${\text{i}}\varepsilon _n \xrightarrow[{}]{}\varepsilon $ from $\mathit{\Sigma} _n^f \left( k \right)$. The Fermi surfaces and the band structure of the diagonalized bands calculated from Eq.(\ref{eq102}) are shown in Fig.\ref{fig1} and Fig.\ref{fig2},  respectively.
Let us introduce the scaling parameter $t_0$ as one eighth of the width of the band-1 for the sake of easy comparison with  usual calculations for a single  band Hubbard model, which corresponds to $V_{\mathvc{k}}  = 0$ in the periodic Anderson model. In figures except for Fig.\ref{fig201},  we rescale all energies (such as $U$ and $E_1 ^\prime  $) by $t_0$ (for example,  shown as $U/t_0$ and $E_1 ^\prime/t_0  $). 
Note that $t_0$ is not equal to $t$, which we set to be unity. The ratio $t_0/t$ is $3 \sim 3.5$ and depends on $U$. When $U/t_0=4.1$, $t_0/t$ is about 3.1. Fig.\ref{fig201} shows the bare density of state without self-energy correction. In Fig.\ref{fig201} energies are rescaled by $t$, not by $t_0$. The bare density of states of $f$-band in the periodic Anderson model $\rho _f \left( \varepsilon  \right)$ and that in single band model $\rho _f^0 \left( \varepsilon  \right)$ are given by
%
%
\begin{eqnarray}
\rho _f \left( \varepsilon  \right) =  - \frac{1}
{\pi }\operatorname{Im} \sum\limits_{\mathvc{k}} {G_0^{fR} \left( {{\mathvc{k}},\varepsilon } \right)}, \label{eq201} \\
\rho _f^0 \left( \varepsilon  \right) =  - \frac{1}
{\pi }\operatorname{Im} \sum\limits_{\mathvc{k}} {G_0^{1R} \left( {{\mathvc{k}},\varepsilon } \right)}, \label{eq202}
\end{eqnarray}
%
%
, respectively. Here, the unperturbed Green's functions 
$G_0^{fR}$ and $G_0^{1R}$ are calculated by analytic continuation ${\text{i}}\varepsilon _n  \to \varepsilon $ from $ G_0^f \left( k \right)$ and $\left( {{\text{i}}\varepsilon _n  - \varepsilon _{\mathvc{k}}^f } \right)^{ - 1} $, respectively. In Fig. \ref{fig201}, the DOS in the periodic Anderson model becomes flat and there exist high and low energy tails in the DOS owing to the hybridization. These facts mean that the width of $f$-band is expanded compared to that in  single band model. The expanded band width  stabilize the Fermi liquid state. Thus our perturbation calculation can be valid owing to the expansion of the band width, even when $U$ exceeds $8t$, which is the band width in the single band model. 
%
%
\begin{figure}
\includegraphics[]{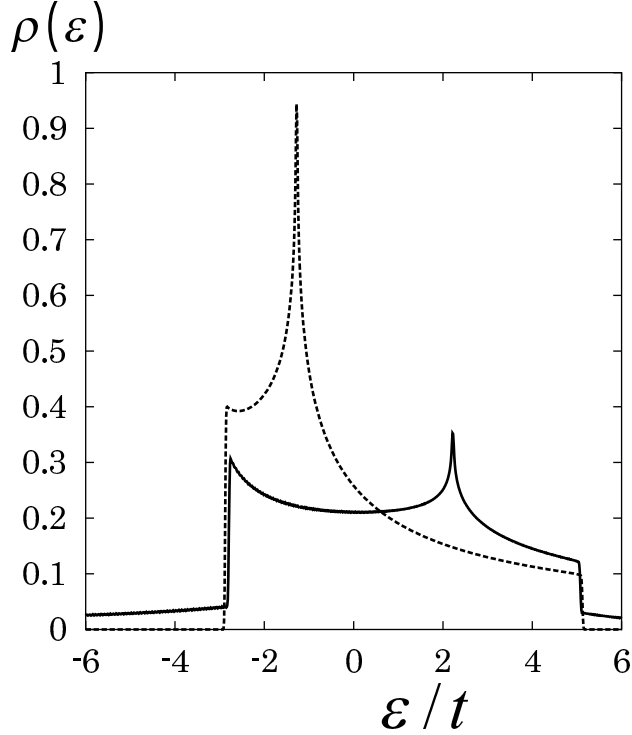}
\caption[]{The DOS for $f$-electrons in periodic Anderson model (solid curve) and in single band model (dashed curve). In periodic Anderson model, the band width is expanded and the DOS is flat compared to those in single band model.}
\label{fig201}
\end{figure}
%
%

We now calculate $T_c$D In our model,    the  Coulomb repulsion works only between $f$-electrons,      so we have only to account of $f$-electron Green's functions. Then,   the   anomalous self-energy $\mathit{\Sigma} _a \left( k \right)$  is given by
\begin{eqnarray}
\mathit{\Sigma} _a \left( k \right) = \mathit{\Sigma} _{{\rm{RPA}}} \left( k \right) + \mathit{\Sigma} _{{\rm{vert}}} \left( k \right),      \label{eq22} \\
\mathit{\Sigma} _{{\rm{RPA}}} \left( k \right) =  - \frac{T}{N}\sum\limits_{k'} {\left[ {U + U^2 \chi _{f0} \left( {k + k'} \right) + 2U^3 \chi _{f0} ^2 \left( {k + k'} \right)} \right]} F\left( {k'} \right),      \label{eq23} \\
\mathit{\Sigma} _{{\rm{vert}}} \left( k \right) =  - U^3 \left( {\frac{T}{N}} \right)^2 \sum\limits_{k', k_1 } {G_0^f \left( {k_1 } \right)\left( {\chi _{f0} \left( {k + k_1 } \right) - \phi _{f0} \left( {k + k_1 } \right)} \right)} G_0^f \left( {k + k_1  - k'} \right)F\left( {k'} \right) \nonumber \\
{} - U^3 \left( {\frac{T}{N}} \right)^2 \sum\limits_{k', k_1 } {G_0^f \left( {k_1 } \right)\left( {\chi _{f0} \left( { - k + k_1 } \right) - \phi _{f0} \left( { - k + k_1 } \right)} \right)} G_0^f \left( { - k + k_1  - k'} \right)F\left( {k'} \right). \label{eq24}
\end{eqnarray}
In superconducting state $G^f \left( k \right)$ and the anomalous Green's function for $f$-electron $F  \left( k \right)$ satisfy Dyson-Gor'kov equations.\cite{bib13} 
%
\begin{eqnarray}
G^f \left( k \right) = G_0^f \left( k \right) + G_0^f \left( k \right)\mathit{\Sigma} _n^f \left( k \right)G^f \left( k \right) + G_0^f \left( k \right)\mathit{\Sigma} _a \left( k \right)F^\dag  \left( k \right) \label{eq106}, \\
F^\dag  \left( k \right) = G_0^f \left( -k \right)\mathit{\Sigma} _n^f \left(- k \right)F^\dag \left( k \right) + G_0^f \left( -k \right)\mathit{\Sigma} _a \left( -k \right)G^f \left( {k} \right) \label{eq107}. 
\end{eqnarray}
%
%
The normal self-energy is calculated in Eq.(\ref{eq12}). 

In the vicinity of $\mathit{T_{c}}$,      $ F\left( k \right)$ can be linearized as
\begin{eqnarray}
F\left( k \right) = \left| {G^f \left( k \right)} \right|^2 \mathit{\Sigma} _a \left( k \right),  \label{eq25} \\
G^f \left( k \right) = G_0^f \left( k \right) + G_0^f \left( k \right)\mathit{\Sigma} _n^f \left( k \right)G^f \left( k \right) \label{eq108}. 
\end{eqnarray} 
Thus,  $\mathit{\Sigma} _a \left( k \right) $ at $T_c$ is determined by the gap equation
%
%
\begin{eqnarray}
\mathit{\Sigma} _a \left( k \right) =  - \frac{T}{N}\sum\limits_{k'} {V_a \left( {k, k'} \right)\left| {G^f \left( {k'} \right)} \right|} ^2 \mathit{\Sigma} _a \left( {k'} \right), \label{eq103}
\end{eqnarray}
where,  
\begin{eqnarray}
V_a \left( {k, k'} \right) = U + U^2 \chi _0 \left( {k + k'} \right) + 2U^3 \chi _0 ^2 \left( {k + k'} \right) \nonumber \\
{} + 2U^3 \frac{T}{N}{\mathop{\rm Re}\nolimits} \sum\limits_{k_1 } {G_0^f \left( {k_1 } \right)G_0^f \left( {k + k_1  - k'} \right)\left[ {\chi _0 \left( {k + k_1 } \right) - \phi _0 \left( {k + k_1 } \right)} \right]}. \label{eq104}
\end{eqnarray}
%
%

\begin{figure}
\includegraphics[]{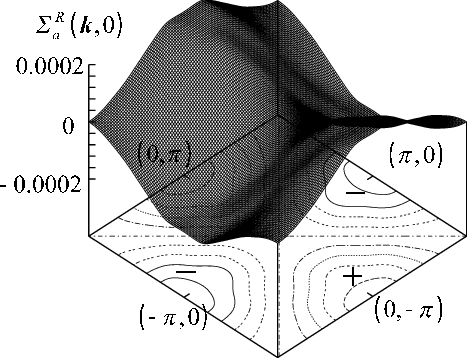}
\caption[]{The $d_\mathrm {x^{2}-y^{2}}$ symmetry of the superconducting gap $\mathit{\Sigma} _a^R \left( {{\mathvc{k}},0} \right)$.}
\label{fig202}
\end{figure}

If we replace the left hand side of Eq.(\ref{eq103}) by $ \lambda \mathit{\Sigma} _a \left( k \right)$, this equation can be considered as an eigenvalue equation with eigenvalue $\lambda$ and eigenvector $\mathit{\Sigma} _a \left( k \right)$. $\mathit{T_{c}}$  is the temperature at which the maximum eigenvalue   reaches to unity. $\mathit{\Sigma} _a \left( k \right)$ represents the superconducting gap symmetry. Among several  gap symmetries, the $d_\mathrm {x^{2}-y^{2}}$ state possesses the maximum eigenvalue. Fig.\ref{fig202} shows the analytic continuation $\mathit{\Sigma} _a^R \left( {{\mathvc{k}},\varepsilon  = 0} \right)$   of the gap function $\mathit{\Sigma} _a \left( k \right)$.

%
%

\begin{figure}
\includegraphics[]{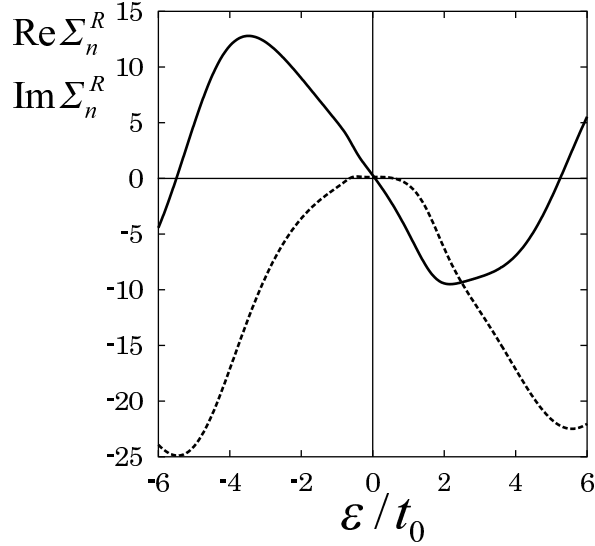}
\caption[]{$\operatorname{Re} \mathit{\Sigma} _n^R $(solid curve) and $\operatorname{Im} \mathit{\Sigma} _n^R $(dashed curve) at $\left( {9\pi /16, 9\pi /16} \right)$ $(U/t_0  = 5.4, T/t_0  = 0.007)$.}
\label{fig101}
\end{figure}

Now we consider the condition in which the perturbation theory in $U$ is valid. We investigate the behavior of retarded normal self-energy $\mathit{\Sigma} _n^R \left( {{\mathvc{k}}, \varepsilon } \right)$. At wavevector ${\mathvc{k}} = {\mathvc{k}}_1  = \left( {9\pi /16, 9\pi /16} \right)$,  which is near the Fermi surfaces of the $f$-band,  $\operatorname{Re} \mathit{\Sigma} _n^R $ and $\operatorname{Im} \mathit{\Sigma} _n^R $ behave as shown in Fig.\ref{fig101}. From Fig.\ref{fig101},  we can see that near $\varepsilon  = 0$,  $\operatorname{Re} \mathit{\Sigma} _n^R  = \alpha \varepsilon $ and $\operatorname{Im} \mathit{\Sigma} _n^R  = \beta \varepsilon ^2 $. These facts show that the retarded normal self-energy behaves as the conventional Fermi-liquid. The perturbation calculation up to third order terms of $U$ is confirmed to be valid.
%
%
%
%

\begin{figure}
\includegraphics[]{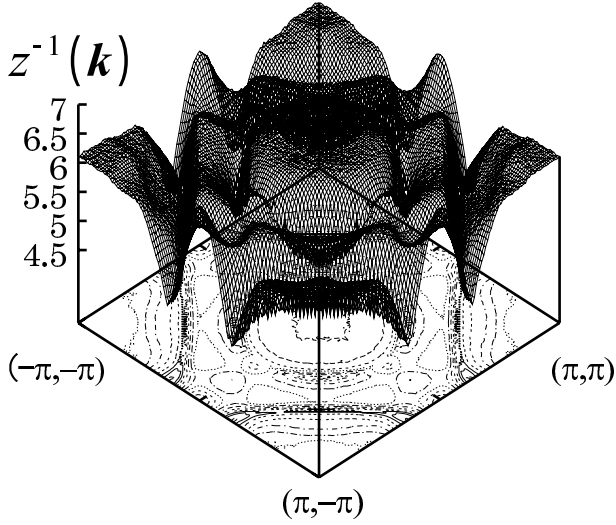}
\caption[]{Quasi-particle mass enhancement factor $z^{-1}\left( {\mathvc{k}} \right)$  $(U/t_0  = 5.4, T/t_0  = 0.007)$. The values of $z^{-1}\left( {\mathvc{k}} \right)$ on the Fermi surfaces of band-1 and band-2 are $4.5 \sim 5.0$.}
\label{fig102}
\end{figure}

Next,  we investigate the  mass enhancement factor for the $f$-band,  $z^{-1}\left( {\mathvc{k}} \right) = 1 - \partial \operatorname{Re} \mathit{\Sigma} _n^R \left( {{\mathvc{k}}, \varepsilon } \right)/\partial \varepsilon $. Fig.\ref{fig102} shows $z^{-1}\left( {\mathvc{k}} \right)$. If the negative contribution of the $U^3$ -term to $z^{-1}\left( {\mathvc{k}} \right)$ is large compared to the $U^2$ -term,  then $z^{-1}\left( {\mathvc{k}} \right)$ becomes near or lower than unity. So the value of $z^{-1}\left( {\mathvc{k}} \right)$ in Fig.\ref{fig102} shows that the contribution of the $U^3$ -term to $z^{-1}\left( {\mathvc{k}} \right)$ is not so large and that perturbation calculation in $U$ is valid. 
%
%
%
%

From Fig.\ref{fig1},  the numbers of filled electrons  in the band-1 and the band-2 are  estimated at 0.34/spin and 0.79/spin,     respectively. In Fig.\ref{fig1},      we can see that the shape of the Fermi surface of the band-1 is preferable to the $d_\mathrm {x^{2}-y^{2}}$ superconductivity,     just like high-$\mathit{T_{c}}$ cuprates. On the other hand, the low value of the electron number in the band-1 (0.34/spin) suppresses AFF,  and lowers the peak of spin susceptibility near $\left( {{\rm{\pi,     \pi }}} \right)$.  The reduced AFF leads to the $d$-wave superconductivity with low $\mathit{T_{c}}$. Surely the band-1 is important for superconductivity,      but we cannot  explain high $\mathit{T_{c}}$ of PuCoGa$_{5}$ if we consider only the  band-1. From these facts,    we consider that the band-2 plays an important role in high $\mathit{T_{c}}$ superconductivity. 

\begin{figure}
\includegraphics[]{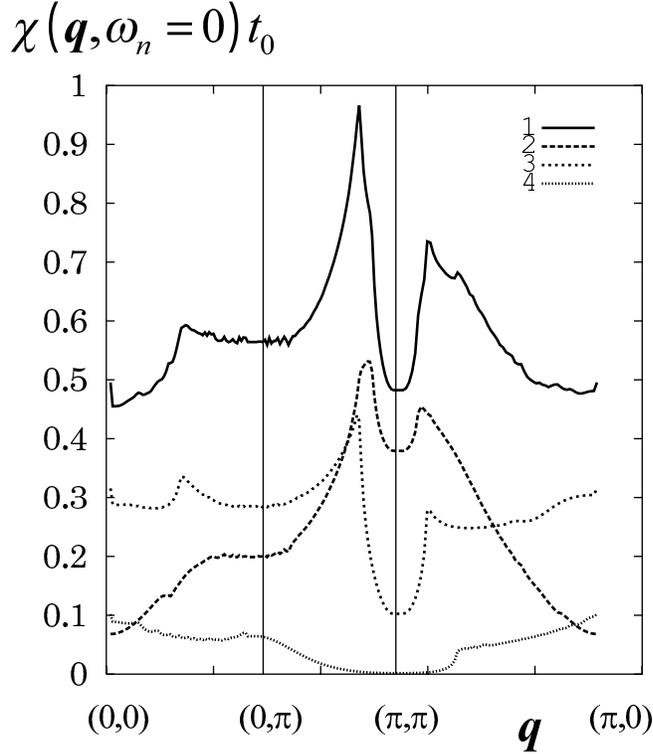}
\caption[]{Spin susceptibility for the single band model and the periodic Anderson model $(U/t_0  = 5.4, T/t_0  = 0.007)$. }
\label{fig3}
\end{figure}
\begin{figure}
\includegraphics[]{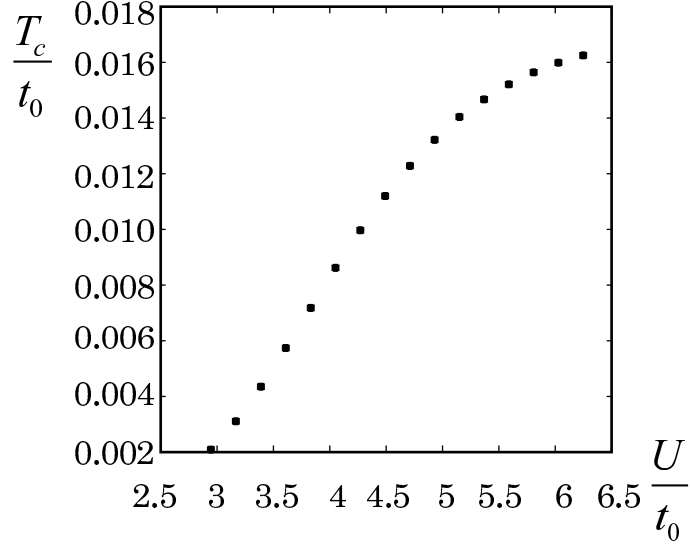}
\caption[]{$\mathit{T_{c}}$ of PuCoGa$_{5}$ calculated by TOPT on the basis of periodic Anderson model. }
\label{fig4}
\end{figure}

Fig.\ref{fig3} shows the spin susceptibility including the self energy correction. Line 1 is spin susceptibility $\chi _{f0} \left( q \right) = \chi _2 \left( q \right) + \chi _3 \left( q \right) + \chi _4 \left( q \right)$ for the periodic Anderson model.  Lines  2,     3 and 4 are $\chi _2 \left( q \right)=\sum\limits_k {c^2 G_1 \left( k \right) c^2 G_1 \left( k+q \right)}$,      $\chi _3 \left( q \right)=2 \operatorname{Re} \sum\limits_k { c^2 G_1 \left( k \right) s^2 G_2 \left( k+q \right)}$ and $\chi _4 \left( q \right)=\sum\limits_k { s^2 G_2 \left( k \right) s^2 G_2 \left( k+q \right)}$,      respectively. There is  a peak near $\left( {{\rm{\pi,     \pi }}} \right)$ for $\chi _2 \left( q \right)$,    which corresponds to  the spin susceptibility for the single band model composed of  the band-1. Moreover,    the exsistence of the coupled contribution between the band-1 and  the band-2 ($\chi _3 \left( q \right)$) causes the higher peak of the spin susceptibility,  which is stronger AFF in our  model,    compared with that in single band Hubbard model. The large susceptibility arises from nesting effects between two bands. The higher $\mathit{T_{c}}$ is originated from this effect. The calculated $\mathit{T_{c}}$ in the periodic Anderson model is shown in Fig \ref{fig4}. The lowest limit of temperature for reliable calculation is approximately $T/t_0=0.002$. From Fig.\ref{fig4} we can see that $\mathit{T_{c}}$ is relatively high even for small values of U,   or weak correlation. We tried to calculate $\mathit{T_{c}}$ in  single band models,      but we could not get any finite value of $\mathit{T_{c}}$ in TOPT because the filling number is far from 0.5/spin. This indicates that $\mathit{T_{c}}$  in the single band model is very low compared with $\mathit{T_{c}}$ in the periodic Anderson model. We can see that in PuCoGa$_{5}$  the $d$-wave superconductivity with high $\mathit{T_{c}}$ is realized even for the modest electron correlation. The modest electron correlation  is consistent with the experimental facts. \cite{bib1}

\section{\label{sec5} Conclusion}
In this paper,      we have explained high $\mathit{T_{c}}$ of PuCoGa$_{5}$ using periodic Anderson model and have shown that the superconductivity in this material is unconventional one  with $d_\mathrm {x^{2}-y^{2}}$ symmetry.  To obtain the results, the following point is important. By considering only the `main' band (the band-1),      it seems impossible to explain high $\mathit{T_{c}}$ because of low filling. In this material,   the existence of  `sub' band  (the band-2) increases much  the density of states at Fermi energy.   Furthermore,   nesting effects between the Fermi surface of the `sub' band and  that of `main' band enhance AFF.  These effects make $\mathit{T_{c}}$ higher even for relatively weak electron correlation. From the band calculation, $W_1 \sim 1\,     {\rm{eV}}\sim 1.16 \times 10^4 {\text{K}}$.\cite{bib10} Since $t_0$ is defined as one eighth of $W_1$, $t_0$ is approximately $ 1.5 \times 10^3{\text{K}}$. Roughly estimated, the superconducting transition temperature    $\mathit{T_{c}}=18.5\mathrm{K}$ corresponds to about $0.012t_0$. From this value of $\mathit{T_{c}}$ and Fig.\ref{fig4}, we can estimate the value of $U/t_0$ at $4 \sim 4.5$. 

This calculation was performed with the computer in Yukawa Institute of theoretical Physics. The authors are grateful to Dr.Y.Nisikawa for valuable discussions.

\end{document}